\documentclass{article}
\usepackage{spconf,amsmath,graphicx}
\usepackage{amsfonts}%
\usepackage{caption}%
\usepackage{graphicx}
\usepackage{subcaption}
\usepackage{booktabs}
\usepackage{multirow}
\usepackage{placeins}
\usepackage{url}
\usepackage{tikz}
\usepackage[nolist,nohyperlinks]{acronym}
\usepackage{amssymb}
\usepackage{pifont}
\usepackage[nolist,nohyperlinks]{acronym}
\usepackage{multirow}
\usepackage[T1]{fontenc}
\usepackage{xcolor}
\usepackage{url}

\acrodef{SV}{Speaker Verification}
\acrodef{ETDNN}{Extended TDNN}
\acrodef{FTDNN}{Factorized TDNN}
\acrodef{SNR}{Signal-to-Noise Ratio}
\acrodef{WADASNR}{Waveform Amplitude Distribution Analysis}
\acrodef{MVN}{Mean-Variance Normalized}
\acrodef{EER}{Equal Error Rate}
\acrodef{minDCF}{Minimum Decision Cost Function}
\acrodef{log mel-FB}{log Mel-filter bank}
\acrodef{DCT}{Discrete Cosine Transform}
\acrodef{CycleGAN}{cycle consistent generative adversarial network}
\acrodef{UEN}{Unsupervised Enhancement Network}
\acrodef{UENs}{Unsupervised Enhancement Networks}
\acrodef{MUSAN}{Music, Speech and Noise}
\acrodef{AMI}{AMI Meeting Corpus}
\acrodef{SITW}{Speakers In The Wild}
\acrodef{RIR}{Room Impulse Response}
\acrodef{Lcyc}{cycle Consistent loss}
\acrodef{Ladv}{adversarial loss}
\acrodef{WPE}{Weighted Prediction Error}
\acrodef{SRE}{Speaker Recognition Evaluation}
\acrodef{VoxSRC}{VoxCeleb Speaker Recognition Challenge}
\acrodef{ASR}{Automatic Speech Recognition}
\acrodef{WER}{Word Error Rate}
\acrodef{SOTA}{state-of-the-art}
\acrodef{PLDA}{Probabilistic Linear Discriminant Analysis}
\acrodef{VAD}{Voice Activity Detection}

\title{Unsupervised Feature Enhancement for speaker verification}
\name{Phani Sankar Nidadavolu,
Saurabh Kataria,
Jes\'us Villalba,
Paola Garc\'ia-Perera,
Najim Dehak\thanks{The research reported here was conducted at the 2019 Frederick Jelinek
Memorial Summer Workshop on Speech and Language Technologies, hosted
at L'\'Ecole de Technologie Sup\'erieure (Montreal, Canada) and sponsored by Johns Hopkins University
with unrestricted gifts from Amazon, Facebook, Google, and Microsoft.}}
\address{Center for Language and Speech Processing, Johns Hopkins University, Baltimore, MD, USA}

\begin{document}
%
\maketitle
\begin{abstract}
The task of making speaker verification systems robust to adverse scenarios remains a challenging and an active area of research. We developed an unsupervised feature enhancement approach in log-filter bank space with the end goal of improving speaker verification performance. We experimented with using both real speech recorded in adverse environments and degraded speech obtained by simulation to train the enhancement systems. The effectiveness of this approach was shown by testing on several real, simulated noisy, and reverberant test sets. The approach yielded significant improvements on both real and simulated sets when data augmentation was not used in speaker verification pipeline. We also experimented with training the x-vector and PLDA systems with enhanced augmented features instead of augmented features and observed better performance on real test conditions (4.2\% relative improvement in minDCF on SRI).

\end{abstract}

\vspace{-3mm}
\section{Introduction}
\label{sec:intro}
\vspace{-2mm}

Speech signals get contaminated by various background noises, reverberation and other unwanted variabilities present during their acquisition. An ideal \ac{SV} system should be robust to any background noises and reverberation effects present. Recently, developing robust \ac{SV} systems has become a very active research area. Several challenges were organized recently such as NIST \ac{SRE} 2019, VOiCES from a Distance Challenge \cite{nandwana2019voices}, and VoxCeleb Speaker Recognition Challenge 2019.

One approach to improve the robustness of \ac{SV} systems is to train them on data created by artificially adding noise to the original training data or simulating reverberant speech. This method, known as data augmentation, has proven to be effective in improving the performance of \ac{SV} systems yielding \ac{SOTA} results on various tasks~\cite{snyder2018x,villalba2019state}. However, such simulation strategies do not take into account the amount and type of degradation the test utterances may have. A recent study on Speaker Diarization on children's speech~\cite{xie2019multi} demonstrates various challenges that x-vector systems face in adverse scenarios. 


In this work, we experimented with an unsupervised single channel wide-band feature enhancement approach to improve the quality of speech features with the end goal of improving the performance of \ac{SV} systems -- a \textit{task-specific} approach. The motivation behind taking an unsupervised approach was to incorporate the knowledge of the target (adverse) domain in the enhancement procedure with the help of some unlabeled training data from that domain. The \ac{UEN} we experimented with was a \ac{CycleGAN} \cite{zhu2017unpaired} trained on \ac{log mel-FB} features.  


Previously, \emph{task-specific} enhancement techniques have been proposed for \ac{ASR} and \ac{SV}.
Denoising approach using \ac{CycleGAN} was proposed by \cite{meng2018cycle} to improve the performance of \ac{ASR} with results reported on several simulated test conditions.
For \ac{SV}, \cite{shon2019voiceid} and \cite{michelsanti2017conditional} have reported improvements on simulated data.


The main contributions of this paper are as follows: 1) to develop a unified \ac{UEN} that serves dual purpose - simultaneous dereverberation and denoising, 2) to test the generalization ability of this network to unseen test conditions, 3) use features extracted from real degraded speech to train the \ac{UEN} and 4) to investigate if the \ac{UEN} approach complements the \ac{SOTA} x-vector system trained with data augmentation. 


Our experimental approach was as follows: we developed an unsupervised enhancement based \ac{SV} pipeline, referred as \ac{UEN}-\ac{SV} system. In \ac{UEN}-\ac{SV} system, we enhance the features of test data (enrollment and evaluation data) before extracting x-vectors. We also experimented with a homogeneous \ac{UEN}-\ac{SV} system where the x-vector network and PLDA were trained on enhanced augmented features instead of augmented features. 

\vspace{-3mm}

\section{Unsupervised Enhancement System}
\label{sec:uen}
\vspace{-2mm}

\subsection{\ac{CycleGAN} Training}
\label{cg_training}
\vspace{-1mm}
The \ac{UEN} in this work is a \ac{CycleGAN} which consists of two generators and two discriminators. The generators map features from one domain \footnote{To be consistent with the notation of CycleGAN, we used the terms clean/source and reverberant/target interchangeably in this paper} to the other. They were trained using a multi-task objective which consists of two loss components- an adversarial loss and a cycle consistent loss. Adversarial loss was responsible for making the generator produce features that appear to be drawn from the opposite domain. Cycle consistency loss additionally constrains the generator to reconstruct original features of the domain from the generated features in opposite domain (achieved by minimizing the $L_1$ distance between original and reconstructed features). The adversarial loss of each generator takes help from a binary classifier, termed as discriminator, coupled to that generator. The task for the discriminator is to classify between original and generated features of a particular domain, achieved by minimizing a least-squares objective \cite{mao2017least}. The adversarial loss then becomes a non saturating loss as shown in~\cite{goodfellow2014generative}. During evaluation, features of degraded speech are enhanced  by mapping them to clean domain using the corresponding generator. More details on the objectives used for training \ac{CycleGAN} can be found in our previous work on domain adaptation~\cite{nidadavolu2019lr,nidadavolu2019cycle}.

\vspace{-5mm}
\subsection{\ac{CycleGAN} Architecture}
\label{cg_archs}
\ac{CycleGAN} generator was a full-convolutional residual network with an encoder-decoder architecture. The encoder consisted of three convolutional layers followed by nine residual blocks. The number of filters in the first three convolutional layers were set to 32, 64 and 128 with strides of 1,2 and 2 respectively. The residual network consisted of two convolutional layers with 128 filters. The decoder network consisted of two deconvolutional layers with strides 2 and filters 64 and 32 respectively followed by a final convolutional layer with stride 1. Instance normalization was used in each layer except the first and last. ReLU activation was used in all layers except the last. The kernel size in all layers was set to 3x3. We used a short cut connection from input of the network to the output (input was added to the output of the last layer which becomes the generator's final output). We trained the generators on \ac{log mel-FB} features. Since, dereverberation is a convolution operation it becomes additive in the log-spectral domain. Hence, the short cut connection disentangles the reverberation effect (which was estimated by the model) from the input. The discriminator had 5 convolutional layers each with a kernel size of 4. The strides of first three and last two layers were set to 2 and 1 respectively. The number of filters in each layer were set to 64, 128, 256, 512 and 1. LeakyReLu with slope 0.2 was used as activation in all layers except the last. More details on the architecture can be found in \cite{nidadavolu2019lr}.

\vspace{-5mm}
\subsection{x-vector Architecture}
\label{xvec_archs}

For the x-vector network in our \ac{SV} pipeline, we experimented with an \ac{ETDNN} architecture~\cite{villalba2019state}. \ac{ETDNN} improves upon TDNN~\cite{snyder2018x} by interleaving dense layers in between the convolution layers. More details on the \ac{ETDNN} network and the pipeline can be found in~\cite{villalba2019state, garcia2019speaker}.

\vspace{-3mm}

\section{Experimental Details}
\label{sec:exp_details}

\subsection{Dataset Details}
\label{dataset_details}

The training of \ac{UEN} network requires access to non-parallel features from clean and reverberant domains which was obtained as follows. The files from the same YouTube video of VoxCeleb1~\cite{nagrani2017voxceleb} and Voxceleb2~\cite{chung2018voxceleb2} were concatenated, denoted as \emph{voxcelebcat}, to obtain longer audio sequences. Since \emph{voxcelebcat} was collected in wild conditions and contained unwanted background noise, additional filtering of files was done based on their \ac{SNR}, similar to the recent LibriTTS~\cite{zen2019libritts} work. We retained only the top 50\% files sorted by their estimated \ac{SNR} value using \ac{WADASNR} algorithm~\cite{kim2008robust}. Thus, we obtained speech from 7104 speakers with duration around 1665 hours. The high SNR signals thus obtained, termed as \emph{voxcelebcat\_wadasnr}, was used as source domain (clean corpus) for training the \ac{UEN}.

Degraded speech from target domain for training the \ac{UEN} was obtained either by simulation or by real recordings collected in adverse conditions.The degraded speech using simulation was obtained by first convolving \emph{voxcelebcat\_wadasnr} with simulated \ac{RIR}\footnote{All \ac{RIR}s are available  at \url{http://www.openslr.org/26}} with RT60 values in the range 0.0-1.0 seconds. Then \textit{noise} from \ac{MUSAN} corpus was artificially added (at \ac{SNR} levels 15,10,5 and 0dB) to the simulated reverberant speech (\textit{speech} and \textit{music} portions from \ac{MUSAN} were not used in the simulation and \textit{noise} was added as \textit{foreground} noise). This corpora, termed as \emph{voxcelebcat\_reverb\_noise}, was used as target domain for training the \ac{UEN}.

The target domain data for \ac{UEN}s trained with degraded speech obtained from real recordings was sampled from training sets of \ac{AMI}~\cite{mccowan2005ami} and Chime5~\cite{barker2015third}. \ac{AMI} was recorded in a setting of 3 different meeting rooms, 180 speakers x 3.5 sessions per speaker. Out of these 180 speakers, 135 speakers were used for training the UEN and 45 for testing. Chime5 corpus was recorded in an indoor uncontrolled setting of kitchen, dining, living room with 80 speakers. Similar to simulated setup, we added \textit{noise} from \ac{MUSAN} to the recordings of \ac{AMI} and Chime5. Addition of noise to reverberant speech followed from our earlier work on domain adaptation~\cite{nidadavolu2019lr} where it was shown that noise addition improves the performance of \ac{CycleGAN} by making the distributions of both domains distinct while also improving the speed of convergence.Clean data, \emph{voxcelebcat\_wadasnr}, remains the same for both simulated and real target domain \ac{UEN} setups. The real target domain has much less speakers (135 from \ac{AMI}) compared to simulated setup (7104). 

To test our \ac{UEN}-\ac{SV} pipeline, we used three different corpora: \ac{SITW} \cite{mclaren2016speakers}, \ac{AMI} and SRI \cite{SRI-Real-Voices}\footnote{This data was recorded by SRI international and was submitted to LDC for publication}. SRI data was recorded in an indoor controlled setting of small/large rooms; controlled backgrounds, 30 speakers $\times$ 2 sessions and 40 hour. SRI data does not have a training portion, we used training corpus from Chime5 (as explained earlier) as target domain for training the \ac{UEN} on real data. To test the effectiveness of the enhancement system, we also tested our \ac{UEN}-\ac{SV} system on reverberant and noisy tests obtained from \ac{SITW} using simulation. We treated SITW as clean corpus. The reverberant copy of SITW, known as SITW \textit{reverb}, was created similar to the training  except that the max value of RT60 for the RIRs used was set to 4.0 seconds (instead of 1.0). We ensured \ac{RIR}s for training and testing simulations were disjoint. We also designed a simulated additive noise testing setup, called \ac{SITW} \textit{noisy}, by adding different types of noise from \ac{MUSAN} corpus and ``background noises'' from CHiME-3 challenge (referred to as \textit{chime3bg}) at different \ac{SNR}s. This resulted in five test SNRs (-5dB, 0dB, 5dB, 10dB, 15dB) and four noise types (\textit{noise}, \textit{music}, \textit{babble}, \textit{chime3bg}). It is ensured that the test \textit{noise} files were disjoint from the ones used for training.

The testing data for AMI and SRI data was split into enrollment and test utterances which
were classified as per their duration. \textit{test>=$n$ sec} and \textit{enroll=$m$ sec} refers to test and enrollment utterances of minimum $n$ and equal to $m$ seconds from the speaker of interest respectively with
$n \in \{0,5,15,30\}$ and $m \in \{5,15,30\}$. The results from all conditions were averaged and reported in this work. 

For x-vector system training, we experimented with training \ac{ETDNN} network without and with data augmentation. The \ac{ETDNN} network without data augmentation was trained on \textit{voxcelebcat\_wadasnr}\footnote{Data preparation and training scripts can be found at: \url{https://github.com/jsalt2019-diadet/jsalt2019-diadet}} (details in \ref{dataset_details}). For training ETDNN with augmentation, data was obtained from \textit{voxcelebcat\_wadasnr} by following the procedure mentioned in ~\cite{snyder2018x}- 3 copies of data were created by artificially adding noise from MUSAN corpus and 1 copy was created by convolving with simulated RIRs and a random subset of these 4 copies along with original \textit{voxcelebcat\_wadasnr} was used as augmented data.

\vspace{-4mm}
\subsection{Training Details}
\label{training_details}
\vspace{-1mm}

\ac{CycleGAN} network was trained on 40-dimensional \ac{log mel-FB} features. Short-time mean centering and energy based \ac{VAD} was applied on the features. Two batches of features were sampled from clean and degraded speech during each training step. Since, the training process was unsupervised both the mini batches were drawn in a completely random fashion with no correspondence between the two batches. Batch size and sequence length were set to 32 and 127 respectively.The model was trained for 50 epochs. Each epoch was set to be complete when one random sample from each of the utterances of clean training corpus has appeared once in that epoch. Adam Optimizer was used with momentum $\beta_1=0.5$. The learning rates for the generators and discriminators were set to 0.0003 and 0.0001 respectively. The learning rates were kept constant for the first 15 epochs and, then, linearly decreased until they reach the minimum learning rate (1e-6). The cycle and adversarial loss weights were set to 2.5 and 1.0 respectively. We trained \ac{ETDNN} using Kaldi for 3 epochs with Natural Gradient Descent optimizer, and multi-GPU periodic model averaging scheme. These x-vector networks were trained with 40-dimensional MFCC features. During evaluation, output \ac{log mel-FB} features of \ac{UEN} were converted to MFCCs by applying \ac{DCT} before forward passing through the x-vector network.

\vspace{-4mm}
\section{Results}
\label{sec:results}
\vspace{-3mm}

In this section, we present the results of \ac{UEN}-\ac{SV} system with and without augmentation applied to SV systems. All the results are reported using metrics \ac{minDCF} and \ac{EER}. 

\begin{table}[htbp]
    \centering
     \caption {Enhancement results on \ac{SITW} and \ac{SITW} \textit{reverb}}
    \begin{tabular}{|l|c|c|c|c|c|c|c|}
        \hline
        & 
        \multicolumn{2}{c|} {\textbf{SITW}} &
        \multicolumn{2}{c|} {\textbf{SITW reverb}}  \\  
        \textbf{ETDNN w/o aug} &EER& minDCF & EER& minDCF  \\
        \hline               
        Baseline \ac{SV} & \textbf{5.23} & 0.340 & 6.78 & 0.460 \\
        \ac{SV} with WPE enh & 5.69 & 0.370 & 6.48 & 0.466 \\
        \textit{sim} \ac{UEN}-\ac{SV} & 5.68 &\textbf{0.323} & \textbf{6.09} & \textbf{0.363}  \\
        \hline
    \end{tabular}

     \label{tab:sitw_reverb_results}
     \vspace{-5mm}
\end{table}

\vspace{-2mm}

\begin{table*}[htbp]
    \centering
     \caption {Enhancement results on SITW \textit{noisy} at various \ac{SNR}s (in dB) (Only DCF values are shown to be concise)}
    \begin{tabular}{|l|c|c|c|c|c|c|c|c|c|c|c|c|c|c|c|c|c|c|}
        \hline
        & 
        \multicolumn{4}{c|} {\textbf{MUSAN noise}} &
        \multicolumn{4}{c|} {\textbf{MUSAN music}} & 
        \multicolumn{4}{c|} {\textbf{MUSAN speech}} & 
        \multicolumn{4}{c|} {\textbf{chime3bg}}  \\ 
        \textbf{ETDNN w/o aug}&10&5&0& -5 &10&5&0& -5 &10&5&0& -5 &10&5&0& -5 \\
        \hline               
         Baseline \ac{SV} & .42 & .50 & .63 & .80 & .39 & .48 & .66 & .87 & .43 & .61 & .89 & 1.0 & .45 & .62 & .92 & .99   \\
        \textit{sim} \ac{UEN}-\ac{SV} & \textbf{.36} & \textbf{.39} & \textbf{.46} & \textbf{.57} & \textbf{.34} & \textbf{.38} & \textbf{.47} & \textbf{.64}  & \textbf{.37} & \textbf{.49} & \textbf{.77} & \textbf{.99} & \textbf{.35} & \textbf{.40} & \textbf{.51} & \textbf{.71}\\
        \hline
    \end{tabular}
     \label{tab:sitw_noisy_results}
\end{table*}

\subsection{\ac{UEN}-\ac{SV} Results on \ac{SITW} and simulated \ac{SITW}}
\label{ssec:simsitw_results}

Table~\ref{tab:sitw_reverb_results} presents the results for \ac{UEN}-\ac{SV} system with \ac{ETDNN} trained without data augmentation on \textit{core}-\textit{core} condition of \ac{SITW} and \ac{SITW} \textit{reverb} test sets. The \ac{UEN} network was trained on simulated \emph{voxcelebcat\_reverb\_noise} data as target domain (details in \ref{dataset_details}), the system was termed as \textit{sim} \ac{UEN}-\ac{SV}. We compared these results with a baseline \ac{SV} system where the test features were not enhanced and a \ac{SV} system where the test datasets were enhanced using the \ac{SOTA} \ac{WPE}~\cite{nakatani2010speech,yoshioka2012generalization} dereverberation algorithm. We obtained 21\% and 22\% relative improvements on \ac{minDCF} of \ac{SITW} \textit{reverb} over baseline \ac{SV} and \ac{SV} with \ac{WPE} enhancement. 

We then tested \textit{sim} \ac{UEN}-\ac{SV} system on \ac{SITW} \textit{noisy} (details in \ref{dataset_details}). Out of the four different testing conditions, only \ac{MUSAN} \textit{noise} was added to the training data of \ac{UEN}. The remaining three conditions (\ac{MUSAN} \textit{speech}, \ac{MUSAN} \textit{music} and \textit{chime3bg}) were not used during the training of \ac{UEN}.The results are presented in Table~\ref{tab:sitw_noisy_results}. \textit{sim} \ac{UEN}-\ac{SV} yielded consistent improvements on all four noise conditions at all \ac{SNR}s. More pronounced improvements were observed at 0dB and -5dB \ac{SNR}s. The results showed that the \ac{UEN} we devised exhibited good dereverberation and denoising capabilities and also good generalization ability to unseen noise conditions (\textit{music}, \textit{speech} and \textit{chime3bg}). 

\vspace{-4mm}
\subsection{\ac{UEN}-\ac{SV} Results on AMI and SRI}
\label{ssec:ami_sri_results}

 Encouraged by the results on \ac{SITW} \textit{reverb} and \ac{SITW} \textit{noisy}, we tested the \ac{UEN}-\ac{SV} system on more challenging evaluation corpora from AMI and SRI. Results are presented in Table~\ref{tab:etdnn_results_on_ami_and_sri}. In addition to \textit{sim} \ac{UEN}-\ac{SV} we also present results of \ac{UEN} system trained using real data as target domain, termed as \textit{real} \ac{UEN}-\ac{SV} and results for \ac{SV} system with PLDA adapted to target domain (details in~\cite{villalbajhu}). The \ac{UEN} system for AMI was trained on the training corpus of AMI. However, the \ac{UEN} system for SRI was trained on Chime5 as target domain data for lack of availability of training set for SRI corpus (details in \ref{dataset_details}). As shown in Table~\ref{tab:etdnn_results_on_ami_and_sri}, both the \textit{real} and \textit{sim} \ac{UEN}-\ac{SV} systems improved in performance compared to the baseline SV system for both the testsets. For AMI, \textit{real} \ac{UEN}-\ac{SV} performed better than \textit{sim} \ac{UEN}-\ac{SV} system even though it was trained on smaller amount of target domain data compared to the \textit{sim} \ac{UEN}. However, the advantage of using real data over simulated dropped when PLDA was adapted to the target domain. For SRI, unlike \ac{AMI}, \textit{sim} \ac{UEN}-\ac{SV} performed better than the \textit{real} \ac{UEN}-\ac{SV}. The difference in domains between SRI (testset) and Chime5 (training set) might have resulted in slighlty poor performance of \textit{real}  \ac{UEN}-\ac{SV} compared to its simulated counterpart. From these experiments we observed that when training and evaluation conditions matched closely in target domain (like in AMI) use of real data over simulated data offered advantage, which justifies our approach for unsupervised enhancement.
 
 \vspace{-3mm}

\begin{table}[htbp]
    \centering
     \caption {\ac{UEN}-\ac{SV} results on AMI and SRI}
    \begin{tabular}{@{}lccccccc@{}}
        \hline
        & 
        \multicolumn{2}{c} {\textbf{AMI}}  &
        \multicolumn{2}{c} {\textbf{SRI}}  \\  
         & EER& minDCF & EER & minDCF \\
        \hline               
        \textbf{ETDNN w/o aug} \\
        $\quad$ Baseline SV & 26.51 & 0.940  & 21.11 & 0.767 \\
        $\quad$ sim UEN-SV  & 20.22 & 0.766  & \textbf{18.63 }& \textbf{0.714} \\
        $\quad$ real UEN-SV & \textbf{19.66} & \textbf{0.726} & 19.92 & 0.732   \\
        \hline
        \textbf{ETDNN w/o aug} \\
        \textbf{and PLDA adapt} \\
        $\quad$ Baseline SV & 22.61 & 0.847  & 19.10 & 0.774 \\
        $\quad$ sim UEN-SV  & 18.57 & \textbf{0.680}  & \textbf{17.26} & \textbf{0.738}\\
        $\quad$ real UEN-SV & \textbf{18.21} & 0.691 & 19.41 & 0.767 \\
        \hline
    \end{tabular}
     \label{tab:etdnn_results_on_ami_and_sri}
     \vspace{-5mm}
\end{table}

\begin{table}[htbp]
    \centering
     \caption {\ac{UEN}-\ac{SV} results on AMI and SRI with x-vector and PLDA augmentation (\textit{aug}, \textit{enh} and \textit{orig} in the table stands for augmented data, enhanced augmented data and original data with no augmentation respectively)}
     \resizebox{\columnwidth}{!}{%
    \begin{tabular}{@{}ccccccc@{}}
        \hline
        \multicolumn{3}{c} {\textbf{Details of data used}}&
        \multicolumn{2}{c} {\textbf{AMI}}&
        \multicolumn{2}{c} {\textbf{SRI}}\\ 
        x-vector &PLDA &test &EER& minDCF &EER& minDCF \\
        \hline              
        \textit{aug} &\textit{aug}& \textit{orig} & \textbf{18.79} & \textbf{0.688} & 14.55 & 0.583\\
        \textit{aug} &\textit{aug}& \textit{enh} & 18.96 & 0.711 & 15.43 & 0.644 \\
        \textit{aug} &\textit{enh}& \textit{enh} & 19.41 & 0.690 & \textbf{14.26} & 0.583 \\
        \textit{enh} &\textit{enh}& \textit{enh} & 18.81 & 0.690 & 14.78 & \textbf{0.559} \\
        \hline
    \end{tabular}%
    }
     \label{tab:ftdnn_results_on_ami}
     \vspace{-0.5cm}
\end{table}

 \subsection{\ac{UEN}-\ac{SV} Results on AMI and SRI with Data Augmentation}
\label{ssec:ami_results}

The results of enhancement on x-vector and PLDA trained with data augmentation are presented in Table~\ref{tab:ftdnn_results_on_ami}. Results without enhancement are presented in first row of Table~\ref{tab:ftdnn_results_on_ami}. Second row presents the results when only test (\textit{eval} and enrollment) data was enhanced.On both AMI and SRI, enhancing only test data deteriorates performance. We then trained a \textit{UEN}-\textit{SV} system where PLDA was trained on enhanced augmented data instead of augmented data (results in row 3). On SRI, this improved the results slightly compared to baseline system (better EER and similar minDCF). On AMI, this still deteriorated the performance compared to baseline. We then trained a homogeneous \ac{UEN}-\ac{SV} system - both x-vector and PLDA were trained on enhanced augmented data instead of augmented data. This system gave better performance in terms of minDCF (4.2\% relative) on SRI compared to baseline while giving almost similar performance as the baseline on AMI. SRI being a challenging dataset than AMI (severe background noise), the results indicate that state-of-the-art SV systems trained on data augmentation can take advantage of enhancement. The \textit{UEN}-\textit{SV} system needs to be trained on enhanced augmentation data to benefit from enhancement.

\vspace{-2mm}

\section{Summary}
\label{sec:summary}
\vspace{-3mm}

We devised an unsupervised feature enhancement network (UEN) with the end goal of improving the performance of x-vector based speaker verification (SV) systems. We first demonstrated the effectiveness of the UEN network to do simultaneous dereverebration and denoising by testing on several simulated noisy and reverberant datasets when data augmentation was not used to train the SV system. We then demonstrated the ability of this approach to make use of real data from \textit{target} domains to do feature enhancement. When SV system was trained on enhanced augmented features, UEN complemented the state-of-the-art SV system when tested on severe degraded conditions (4.2\% relative improvement in minDCF on SRI) while giving similar performance as the SOTA system on mildly degraded conditions (AMI).


\clearpage

\bibliographystyle{IEEEbib}
\bibliography{strings,refs}

\end{document}